\documentclass[twocolumn,aps,floats,superscriptaddress,showpacs]{revtex4}
\usepackage{graphicx}
\usepackage{graphicx}

\def\ef {$E_F$}
\def\tas {$1T$-TaS$_2$}
\def\tase {$1T$-TaSe$_2$}
\def\dz2  {$d_{z^{2}}$}

\begin{document}

\bibliographystyle {plain}
\vspace{0.5cm}

\title{Spectroscopic signatures of a bandwidth-controlled
Mott transition at the surface of \em 1T\rm-TaSe$_2$}
\author{L. Perfetti}
\affiliation{Institut de Physique des Nanostructures, Ecole
Polytechnique F\'ed\'erale (EPFL), CH-1015 Lausanne, Switzerland}
\author{A. Georges} \author{S. Florens}
\affiliation{Laboratoire de Physique Th\'eorique de l'Ecole
Normale Sup\'erieure, 24 rue Lhomond, F-75231 Paris Cedex 05, France}
\author{S. Biermann}
\affiliation{Laboratoire de Physique des Solides,
Universit\'e Paris-Sud, Bat. 510, 91405 Orsay, France}
\author{S. Mitrovic} \author{H. Berger}
\affiliation{Institut de Physique des Nanostructures, Ecole
Polytechnique F\'ed\'erale (EPFL), CH-1015 Lausanne, Switzerland}
\author{ Y. Tomm}
\affiliation{Department of Solar Energetics, Hahn-Meitner Institut,
Glienicker Str. 100, D-14109 Berlin, Germany}
\author{H. H\"ochst}
\affiliation{Synchrotron Radiation Center, University of
Wisconsin, Stoughton, WI 53589-3097, USA}
\author{M. Grioni}
\affiliation{Institut de Physique des Nanostructures, Ecole
Polytechnique F\'ed\'erale (EPFL), CH-1015 Lausanne, Switzerland}
\begin{abstract}
\vspace{1cm}
High-resolution angle-resolved photoemission (ARPES) data show that a
metal-insulator Mott transition occurs at
the surface of the quasi-two dimensional compound \tase. The
transition is driven by the narrowing of the Ta \em 5d\rm~ band induced by a
temperature-dependent modulation of the atomic positions. A dynamical
mean-field theory calculation of the spectral function of the
half-filled Hubbard model captures the main qualitative
feature of the data, namely
the rapid transfer of spectral weight from the observed
quasiparticle peak at the Fermi surface to the Hubbard bands, as
the correlation gap opens up.
\end{abstract}
\date{\today}

\maketitle

Electronic correlations can modify the electronic structure of solids
not only quantitatively, but also qualitatively,
inducing new broken-symmetry phases which exhibit charge, spin or
orbital-order, and more exotic
states in low dimensions. One of the most notable consequences
of electronic correlations is the much studied
metal-insulator (M-I) Mott transition \cite{mott,imada}.
Recently, new theoretical approaches have considerably
extended our understanding of this fundamental problem
\cite{dmft-review,imada}.

Many physical properties indirectly reflect the
dramatic rearrangement of the electronic structure at the transition.
Photoelectron spectroscopy, which probes the
single-particle spectral function, can provide a
direct view of such changes \cite{fujimori,allen1,shin}.
However, comparing samples with different compositions
faces materials problems like stoichiometry, defects, and disorder. A
quantitative analysis is further complicated by the known surface
sensitivity of the technique \cite{maiti}\cite{suga}.
An ideal experiment would record
the energy and momentum-dependent spectrum, while tuning the crucial ($W/U$)
parameter ($U$ is the on-site Coulomb correlation energy; $W$ is the
bandwidth) \em in the same single crystal sample\rm. Remarkably, it is
possible to approach this ideal situation exploiting the
occurrence of modulated structures (charge-density-waves; CDWs)
in appropriate low-dimensional systems.
There, the lattice distortion modulates the transfer
integrals and therefore modifies the bandwidth. In materials that are
close enough to a Mott transition, the reduced bandwidth
may lead to an instability. There are strong
indications for this scenario in the layered chalcogenide \tas,
which presents a sharp order-of-magnitude increase of the resistivity
at T=180 K \cite{wilson,tosatti}, with a strong rearrangement of the
electronic states \cite{himpsel,smith,manzke,claessen,zwick}. However,
the complex phase diagram of the CDW in \tas~
affects the electronic transition, which cannot be considered as a
typical Mott transition.

Isostructural and isoelectronic \tase~ exhibits a
similar CDW, but only one phase below T$_C$=475 K.  Its
electrical resistivity remains metallic - albeit rather large -
to very low temperatures \cite{wilson}, suggesting that the
Se compound lies further
from the instability than the S analog. Nevertheless, a
transition could still occur at the crystal surface, where
the $U/W$ ratio is expected to be larger as a result of smaller screening and
coordination. The
surface sensitivity of ARPES is ideal to investigate
such an inhomogeneous state. In this Letter we present
evidence for a Mott transition at the surface of
\tase . High-resolution temperature-dependent ARPES shows
for the first time the disappearance of the coherent
quasiparticle signatures at the Fermi surface and the opening of a
correlation gap. These results are  qualitatively well described
by a dynamical
mean-field (DMFT) calculation, and provide new insight into the
spectral properties of the MI transition.

\tase~ has a layered structure, with the
$d^{1}$ Ta atoms in a
distorted octahedral environment. Adjacent
layers interact weakly through van der Waals gaps, and all physical
properties exhibit a strong anisotropy. A
threefold CDW develops below $T_C=475 K$, with a commensurate
$\sqrt13\times\sqrt13$ superstructure, analogous to the much
studied low-temperature CDW phase of \tas \cite{wilson}.
In real space the CDW
corresponds to a modulation of the atomic positions, within a 13 Ta atoms
unit, in the so-called `star-of-David' configuration.
Extended Huckel calculations \cite{whangbo} suggest that the CDW splits
the Ta \em d\rm~ conduction band into subbands which contain
a total of 13 electrons per unit cell.
Two subbands, carrying 6 electrons each, are
filled and lie below the Fermi level.
The Fermi surface is formed by a half-filled subband carrying
the 13th electron. The opening of a correlation gap in this
subband is responsible for the resistivity jump in
(bulk) \tas \cite{tosatti}.

We performed high-resolution ARPES measurements in Lausanne and
at the PGM beamline of the SRC, University of Wisconsin.
The energy
and momentum resolution were $\Delta$E=10 meV and
$\Delta k = \pm 0.02 $\AA$^{-1}$, and the Fermi level location was determined
with an accuracy of $\pm$ 1 meV by measuring the metallic edge of a
polycrystalline gold reference. Single crystal samples
grown by the usual iodine transport
technique were characterized by Laue diffraction and resistivity
measurements, which confirmed the assignment to the \em 1T\rm~ polytype. They
were mounted on the tip of a
closed-cycle refrigerator and cleaved at a base
pressure of $1\times 10^{-10}$ torr. We did not observe any sign of
surface degradation or contamination over a typical 8 hours run.

Figure 1 shows ARPES intensity maps (h$\nu$ = 21 eV) measured at 300 K and 70 K
along the high-symmetry $\Gamma$M direction of the hexagonal Brillouin zone
(BZ). A higher photon energy (h$\nu$ = 50 eV) yields similar results. The maps correspond to the same CDW phase, as
confirmed by low-energy electron diffraction (LEED), but exhibit
remarkable differences. At 300 K the narrow topmost Ta \em d\rm~ subband
crosses the Fermi level at $k_{F_{1,2}}$
$\sim \pm$ {\small 1/4} $\Gamma M$, in good agreement with band
structure calculations \cite{aebi}. The filled CDW subbands
are visible at $\sim$ 0.3 eV and, with lower intensity
at larger binding energies ($\sim$ 0.8 eV) and momenta. The overall
dispersion of the Ta \em d\rm~ band is influenced by the CDW superlattice, as
will be discussed elsewhere. The parabolic
band with a maximum at $\Gamma$ and $\sim$0.5 eV is a Se \em p\rm~ band.
At 70 K the Ta \em d\rm~ spectral weight is narrower and clearly
removed from \ef, and a gap has appeared.

\begin{figure}[htbp]
\begin{center}
\includegraphics[width=9cm]{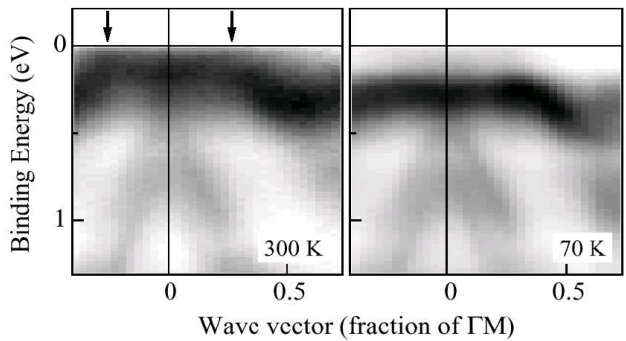}
\end{center}
\caption{
ARPES intensity maps of \tase~ at T=300 K (left) and T=70 K (right)
measured along the $\Gamma$M high-symmetry direction (h$\nu$=21 eV).
The arrows
mark Fermi level crossings by the Ta `d' band at 300 K.
}
\label{fig1}
\end{figure}

Bulk sensitive properties and surface sensitive
LEED data rule out a structural phase transition between 300 K and 70 K.
The large spectral changes are therefore the
consequence of an electronic surface transition. Signatures of a
surface gap in \tase~ were previously observed
at 70 K by scanning tunneling spectroscopy (STS) \cite{hasegawa}.
We characterized
this transition by temperature-dependent measurements of
the shallow Ta $4f$ core levels, which exhibit a CDW-induced
fine structure  \cite{himpsel,scarfe}. We find (not shown) that
between 300 K and 70 K the CDW-split
components sharpen, and their energy separation increases by 40 meV,
in agreement with lower-resolution data \cite{hughes}.
We conclude that the CDW amplitude and the corresponding lattice
distortion are larger at the lower temperature.

The increased distortion reduces the overlap
between the `cluster orbitals' which form the basis of the band
structure in the CDW phase. Calculations which
explicitely account for this effect are not available, but the ($W/U$)
ratio is certainly reduced, possibly below the critical value for the
M-I transition. This is confirmed by an
inspection of the ARPES signal 
at $k=k_{F_{1}}$
(Fig. 2a), which reveals a sudden loss of intensity near \ef~ below $\sim$260 K, and the
appearence of a strong signal centered at 0.26 eV. All spectra were normalized to the same integrated area.
The intensity at \ef~
(Fig. 2b) exhibits a sharp step around 260 K, and a further linear
decrease at lower temperature. The
intensity and the spectral lineshape (not shown) are recovered
upon heating, but
only at higher temperatures. The large hysteresis ($\sim 80 
K$) suggests a
first-order transition \cite{hor}.

\begin{figure}[htbp]
\begin{center}
\includegraphics[width=9cm]{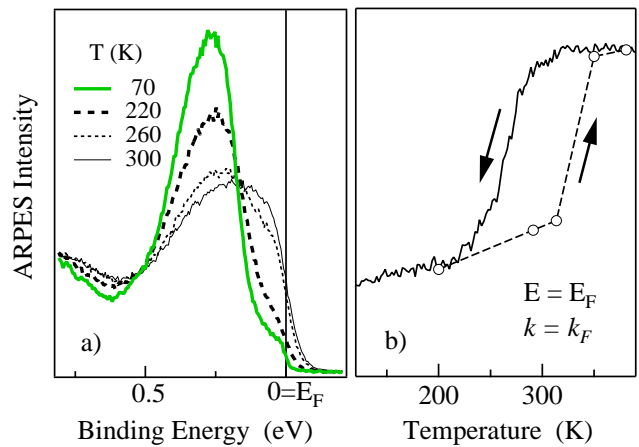}
\end{center}
\caption{
a) ARPES spectra measured at $k=k_{F_{1}}$ between 300 K and 70 K;
b) Temperature dependence of the ARPES signal
at the Fermi surface, showing a sharp break, and a large hysteresis.
}
\label{fig2}
\end{figure}

In Fig. 3, following common practice in
ARPES work on the cuprates, we symmetrized the spectra of Fig. 2a
around \ef. This procedure removes the perturbing effect
of the Fermi distribution on the intrinsic temperature dependence
of the spectral
function $A(k_{F}, \omega, T)$. The 300 K spectrum exhibits a broad
($\sim$1 eV)
incoherent background and a weak quasiparticle (QP) feature at \ef.
The QP signal disappears at lower temperature, and spectral
weight is transferred to the lower and - as inferred by symmetry -
upper sidebands, representing the lower (LHB) and upper (UHB) Hubbard
subbands. The integrated intensity is conserved. We emphasize that
the width ($\sim$100 meV) of the central peak
is much larger than expected for a coherent quasi-particle at the
Fermi surface of a `good' metal. Clearly, in the 220-260 K temperature range,
the corresponding excitations are heavily scattered, and their lifetime
is short. This `bad metal' character is consistent with the broad
maximum and large value of the electrical resistivity of bulk \tase~
at 250-300 K \cite{wilson}.

\begin{figure}[htbp]
\begin{center}
\includegraphics[width=9cm]{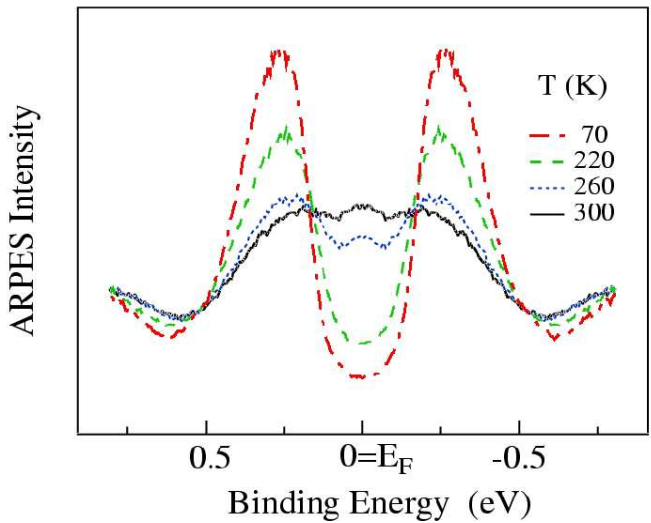}
\end{center}
\caption{
Temperature-dependent ARPES spectral function at $k=k_{F_{1}}$
(h$\nu$=21 eV).  The
spectra have been obtained from the raw spectra I(E) by symmetrization
around \ef: I*(E)=I(E)+I(-E).
}
\label{fig3}
\end{figure}

The spectra of Fig. 3 are qualitatively consistent with the changes expected
at the Mott transition. In order to substantiate this, we have
calculated $A(k_F,\omega;T)$
for a one-band Hubbard model at half-filling, within the dynamical mean-field
theory (DMFT) framework \cite{dmft-review}.
The ``iterated perturbation theory'' approximation
\cite{georges-kotliar} was used, and checks were made using Quantum Monte Carlo
and the maximum entropy method.
The Coulomb term $U$ was set at 0.52 eV, equal to the energy
separation between the LHB and UHB features in Fig. 3.
A semi-circular density of states was used, with a bandwith $W$ assumed
to depend linearly on temperature between 300 K ($W$=0.50 eV) and
70 K ($W$=0.36 eV).
These values are only indicative, and are not the result
of a specific attempt to find an optimum fit to the data, but we note that
the overall magnitude of $W$ is consistent with the dispersions observed
in Fig.1.

The high-temperature (300 K, 260 K)
calculated spectra (Fig. 4) correspond to a correlated metal in the incoherent
regime, with a broad low-energy peak and two intense Hubbard
sidebands. The central peak is strongly reduced at 220 K, and at 70 K it has
disappeared completely, leaving two sharp features centered at $\pm U$/2 and
separated by a real gap.
The overall shape of the spectra is in good qualitative agreement
with the data,
as well as the dramatic transfer of spectral weight that takes place
between the central peak and the Hubbard bands as the temperature is lowered.
We observe that in both theory and experiment
the spectral weight accumulated in the insulator near the maximum of
the HB sidebands comes mainly from the QP central peak, but that a small fraction also comes from
energies above 0.5 eV. This results in two energies at which all
spectra approximately
cross. The data provide the first direct momentum-resolved observation
of two of the key
predictions of DMFT regarding the one-particle spectrum through the Mott
transition, namely the three-peak structure in the metallic state
\cite{georges-kotliar}
and the large transfers of spectral weight from the
metal to the insulator
\cite{georges-krauth,rozenberg,allen2}

\begin{figure}[htbp]
\begin{center}
\includegraphics[width=7cm]{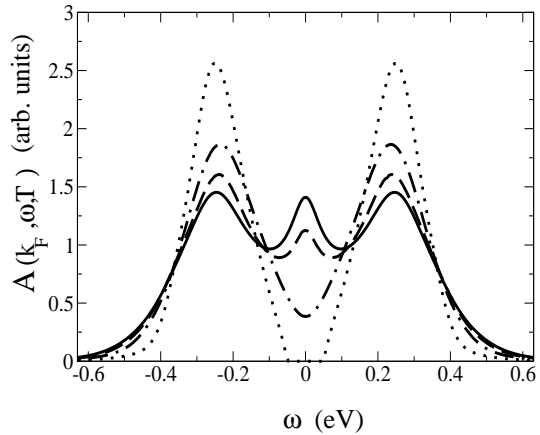}
\end{center}
\caption{
Temperature-dependent spectral function $A(k_{F},\omega)$ of the half-filled
Hubbard model calculated within DMFT for the temperatures of Fig.
4. U was fixed at U=0.52 eV, and the bandwidth is chosen as: W=0.50 eV
at 300 K (solid); 0.48 eV at 260K (dash); 0.44 eV at 220K (dash-dot); and 0.36 eV at 70 K (dot).
}
\label{fig4}
\end{figure}

The comparison of theory and experiment also reveals some differences.
Unlike the calculated spectra, the intensity at \ef~ is never completely
suppressed in the experimental spectra. This signal could
originate
from the underlying metallic bulk, or from surface inhomogeneities (metallic
`patches') observed by STS \cite{hasegawa}. It could also
indicate that the tails of the LHB and UHB overlap slightly, a
passibility which was considered in a different context \cite{thouless}.
Another quantitative difference is the sharper separation between
the QP peak and Hubbard bands in the calculation.
This 
is a known feature of DMFT which is likely to be weakened as
dimensionality is lowered. Recent work in particular \cite{florens1}
suggests that long-wavelength charge modes partially fill the
preformed gap.

From a theoretical standpoint, the main discrepancy concerns
the value of the critical temperature at which the first-order
metal insulator transition is observed. Indeed, with the observed value
of $U$, the
simple one-band Hubbard model treated within DMFT would have a first-order
transition at T$\simeq U/80 \simeq 90 $~K, a factor of three smaller
than observed experimentally. The trajectory in the (T,($W/U$)) space
used in the theoretical calculation does not intersect the first-order
transition line, so that theory would interpret the spectral changes as
due to a rapid crossover between a bad metal and a Mott insulator.
On the other hand the experimental observation of a hysteresis does
suggest a real transition.
In a purely electronic model, it is known that orbital degeneracy does
lead to increased $T_c$ values \cite{florens2}.
However, DMFT model calculations including the whole $d$ manifold,
show a significant increase in $T_c$ only when the
first subband is much closer to the Fermi level than in the experiment.
Thus, it is unlikely that orbital degeneracy could explain the increased
$T_c$.
The inadequacy of a purely electronic model is also suggested by the large
width of the low-energy QP. In a purely electronic model,
this can only be accounted for
if the temperature is significantly larger than the critical temperature
for the metal-insulator transition.

The observation of a broad peak and a true transition
strongly suggests that the coupling to lattice degrees of freedom plays an
important role, as in fact expected in a CDW compound. Indeed, it has been
shown in a simple model that the coupling to the lattice can lead to an
increase in $T_C$ \cite{majumdar}. Also, it is possible that the
observed hysteresis results from differences in the pinning of the CDW
upon heating and cooling. Obviously, further investigation - both theoretical
and experimental - is required to clarify these issues.

In summary, we presented momentum-resolved high-resolution ARPES data
which illustrate the spectral consequences of a bandwidth-controlled
surface Mott transition in \tase.  The transition from a bad-metal,
characterized by a largely incoherent spectrum, to a correlated
insulator is qualitatively captured by a DMFT calculation for the
half-filled Hubbard model. Quantitative differences between theory and
experiment suggest that the model should be extended to include the
coupling to lattice degrees of freedom, in order to provide a more accurate
description of electronic transitions in such CDW materials.

We acknowledge R. Gaal for the resistivity measurements,
correspondence with P. Aebi, E. Canadell, F. del Dongo,
and the support of G. Margaritondo.
This work has been supported by the Swiss NSF through the MaNEP NCCR.
A.G is grateful to F.Mila for discussions and hospitality
at UNIL-Lausanne. A.G, S.F and S.B are most grateful to KITP-UCSB for the warm
hospitality during the final stage of this work, where it
was supported in part by NSF under Grant PHY99-07949, and also acknowledge
computing time at IDRIS-CNRS under project 011393.
The Synchrotron Radiation Center, University of
Wisconsin-Madison, is supported by NSF under Award No. DMR-0084402.

\end{document}